\documentclass{PoS}

\title{Probing BSM physics with electron-proton colliders}

\ShortTitle{Probing BSM physics with electron-proton colliders}

\author{David Curtin\\
        Maryland Center for Fundamental Physics, Department of Physics, University of Maryland, College Park, MD 20742-4111 USA \\
        E-mail: \email{david.r.curtin@gmail.com}}

\author{Kaustubh Deshpande\\
         Maryland Center for Fundamental Physics, Department of Physics, University of Maryland, College Park, MD 20742-4111 USA\\
        E-mail: \email{kau11desh@gmail.com}}

\author{Oliver Fischer\\
        Institute for Nuclear Physics (IKP), Karlsruhe Institute of Technology, Hermann-von-Helmholtz-Platz 1, D-76344 Eggenstein-Leopoldshafen, Germany\\
        E-mail: \email{oliver.fischer@kit.edu}}

\author{\speaker{Jos\'e Zurita} \\%
        Institute for Nuclear Physics (IKP), Karlsruhe Institute of Technology, Hermann-von-Helmholtz-Platz 1, D-76344 Eggenstein-Leopoldshafen, Germany \\
Institute for Theoretical Particle Physics (TTP), Karlsruhe Institute of Technology, Engesserstrae 7, D-76128 Karlsruhe, Germany\\
        E-mail: \email{jose.zurita@kit.edu}}

\abstract{In this talk I will illustrate with two examples (Higgsino dark matter and Exotic Higgs decays) how electron-proton colliders present unique opportunities to probe BSM scenarios where proton-proton colliders fall short due to the experimental difficulties in reconstructing the signal due to the large hadronic backgrounds. The \emph{leit-motiv} of these examples are long-lived particles (LLPs), which have received recently a lot of attention from both the experimental and theoretical communities. We find that the proposed $e^-p$ colliders can be competitive against their more energetic $pp$ incarnations for lifetimes between a millimeter and a micron, depending on the physics scenario under consideration.  }

\FullConference{XXVI International Workshop on Deep-Inelastic Scattering and Related Subjects (DIS2018)\\
		16-20 April 2018\\
		Kobe, Japan}

\begin{document}

\section{Introduction}
\label{sec:intro}

While the lore is that electron-proton ($e^- p$) colliders provide a unique insight into the proton structure, their capabilities to probe Beyond Standard Model (BSM) physics have not been fully explored. It is thus a fair question to ask where the opportunities for discoveries lie. The physics program of the LHC have pushed our naive notion of naturalness as a guiding principle, and the forethought discoveries have not materialized so far.

In that light, $e^- p$ colliders are uniquely suited to look for New Physics that can not be seen at their $pp$ counterparts since it operates in \emph{stealth mode}: compressed spectra, hadronic resonances, tiny couplings are extremely hard at proton-proton colliders. Electron-proton colliders offer a much cleaner environment in terms of hadronic backgrounds (comparable to an $e^+e^-$ machine), albeit with a center of mass energy smaller than (larger than) a $pp$ ($e^+ e^-$) apparatus by a factor of $\sqrt{E_p/E_e}$. This strongly motivates the consideration of signatures which are difficult at a $pp$ collider and that would suffer from the limited center of mass energy of an $e^+ e^-$ machine.

 A nice example of such a stealth signature is provided by Long-Lived Particles (LLPs), which in this talk we will define as BSM states with macroscopic lifetimes.
 From a theoretical perspective, LLPs not only do already exist in the SM, but such particles also appear ubiquitiously in models trying to address fundamental questions such as the hierarchy problem, the origin of neutrino masses, electroweak baryogenesis and the nature of dark matter (for a review of these scenarios see~\cite{Curtin:2018mvb}).

In this talk I will concentrate on two such LLP scenarios. In the first one I will consider a singlet-doublet model of dark matter (reminiscent of the Bino-Higgsino scenario of the MSSM), where the splittings among the doublet components are of a few hundred MeV. At electron-proton colliders this will lead to a novel signature: \emph{displaced pions}. In the second example I will consider Exotic Higgs decays, where the SM Higgs mixes with an additional scalar. The current Higgs data forces this mixing to be small, thus guaranteeing the long-livedness of the new scalar, which gives displaced vertexes in the final state. Both scenarios have been analyzed in detail in~\cite{Curtin:2017bxr} and we refer the reader to that work for further details.

\section{Higgsino dark matter}

Several cosmological and astrophysical phenomena (anisotropies in the Cosmic Microwave Background, rotation curves of galaxies, graviational lensing, formation of large scale structures, etc) can be explained if a new type of matter if introduced. From the purely particle-physics perspective, we do know that dark matter is most-likely neutral under electromagnetism and chromodynamics. We also have a precise measurement of its left-over abundance (relic density). Joining these two pieces of information, one has that the right relic can be reproduced by particles with SU(2) weak couplings ($\alpha \sim 0.01$) and masses of the order of the electroweak scale, $m \sim 100$ GeV.

Hence a simple approach to dark matter is to consider it a single multiplet of SU(2) such that a component is electrically neutral. The classification of those models was carried out in ref.~\cite{Cirelli:2005uq}, and it turns our that all these models are within the reach of the LHC and FCC (see e.g ~\cite{Cirelli:2014dsa,Low:2014cba,Ostdiek:2015aga,Ismail:2016zby}), with the exception of a Majorana fermion, weak doublet, which corresponds to the \emph{pure Higgsino} scenario of the MSSM. In principle such a setup would be testable via disappearing tracks at the FCC~\cite{Mahbubani:2017gjh,Fukuda:2017jmk}.

The model features a weak doublet with a mass term $\mu$ and a heavier singlet with a mass term $M_1$, with $\mu < M_1$. For concreteness we work in the framework of the MSSM, where the free parameters are $M_1, \mu$ and $\tan \beta$, and the Winos are decoupled. Hence in the low energy regime we only have one charged and two neutral states. We thus can describe our model in terms of  $m_{\chi_1^0}$, $\Delta_0=m_{\chi_2^0}-m_{\chi_1^0}$ and  $\Delta_+=m_{\chi^+}-m_{\chi_1^0}$. In the limit where $m_Z, \mu << M_1$ and neglecting higher order terms in $\mu/M_1$, $m_Z / M_1$ we have
\begin{equation}
\Delta_+ = \Delta_{ \rm{1-loop} } + 96~\rm{MeV} (1 \mp s_{2 \beta} ) \Biggl( \frac{10 ~\rm{TeV}}{M_1}  \Biggr) 
\, , \qquad \Delta_0 = 192~\rm{MeV}   \Biggl( \frac{10 ~\rm{TeV}}{M_1}   \Biggr) 
\end{equation}
where $\Delta_{\rm 1-loop}$ is the contribution to the charged-neutral splitting coming from loops of $W,\gamma$ after electroweak symmetry is broken~\cite{Thomas:1998wy}, and it accounts for 200-340 MeV in the $\mu \in [200-1100]$ GeV range (or $c \tau \in [19.6 - 6.7]$ mm) in the limit where $M_1$ is decoupled. Given these value for $\Delta_+$, the chargino decays via $\chi^{\pm} \to \pi^{\pm} \chi_1^0$ with a branching ratio very close to unity. The upper value of 1.1 TeV for $\mu$ corresponds to the saturation of the relic abundance. Finally, since the Z couples to $\chi_1^0-\chi_2^0$ to avoid inelastic scattering in direct detection experiments~\cite{TuckerSmith:2002af,TuckerSmith:2004jv}
 one requires $\Delta_0 \gtrsim 100$ KeV which corresponds to an upper bound on $M_1 < 20 $ PeV. Since $\Delta_0 << \Delta_+$ if $M_1 >> \mu$ we can display our results in terms of the chargino mass and the lifetime $c \tau$. To compute the chargino decay widths we have proceed as in~\cite{Chen:1996ap,Chen:1999yf}.
 
\section{Displaced pions at $e^- p$ colliders}
 
\begin{figure}[tp]
\begin{center}
\includegraphics[width=3.6cm]{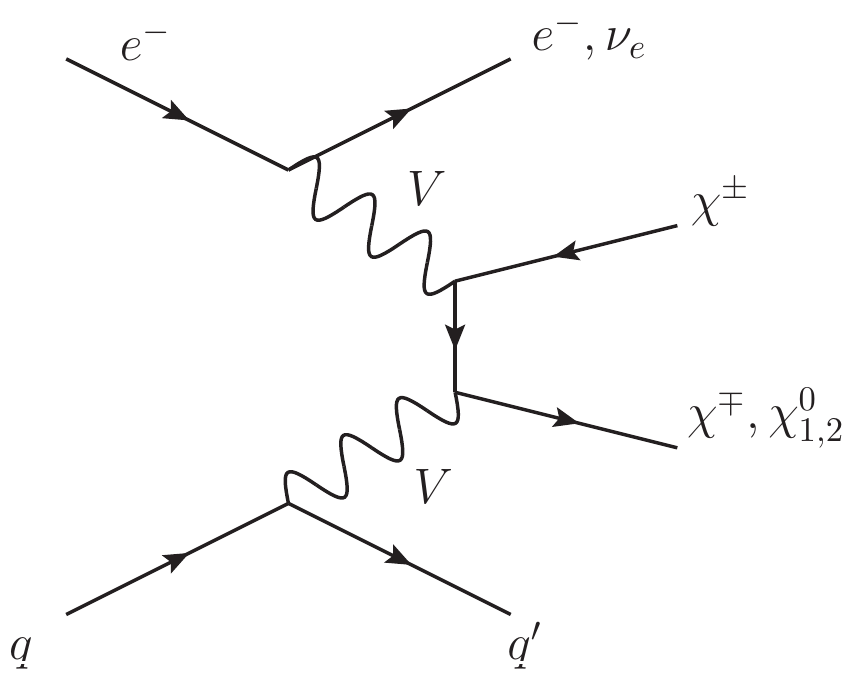}
\hspace{2mm}
\includegraphics[width=4.5cm]{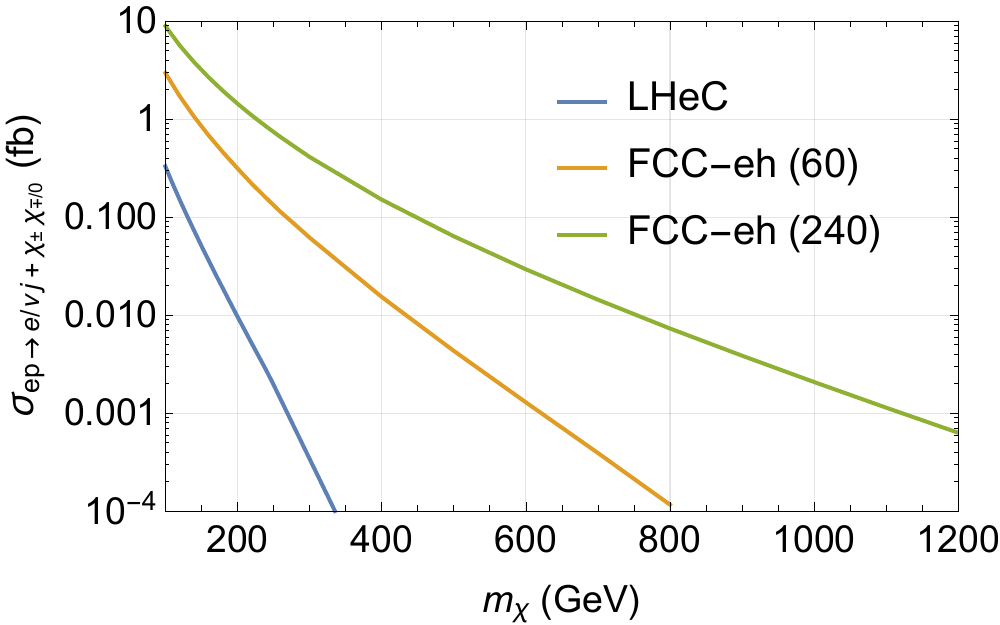}
\hspace{2mm}
\includegraphics[width=5.9cm]{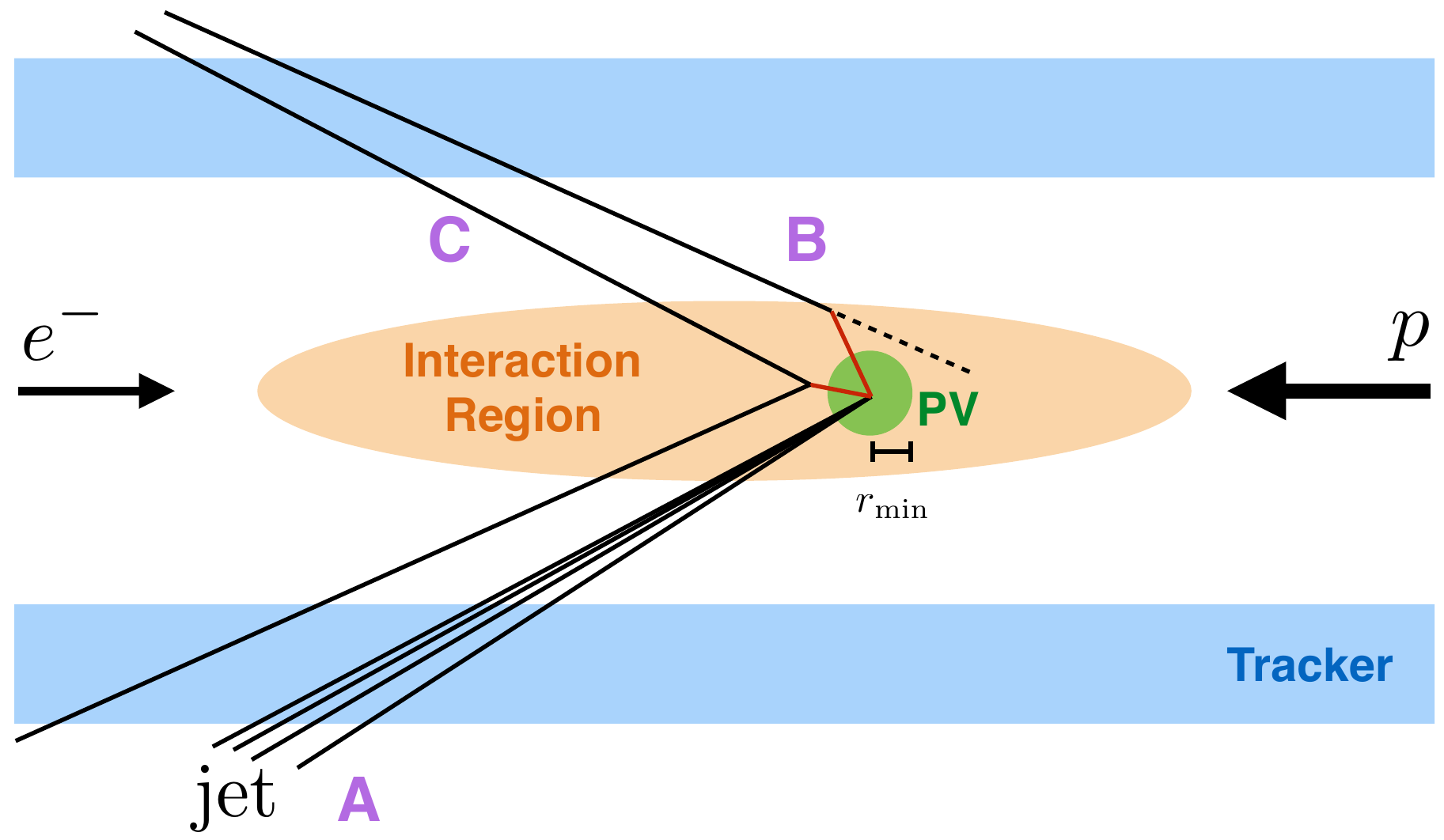}
\end{center}
\vspace{-0.5cm} 
\caption{\label{fig:1}{ Left: Feynman diagram for Higgsino production at an $e^-p$ collider. Center: Production cross section for Higgsino pairs at different $e^-p$ colliders (see main text for details). Right: Sketch of our final states. The remnant of the proton leaves a forward jet (A) used to tag the primary vertex. The chargino is produced and decay in (B) into a pion (black solid) and a neutralino (not shown). The pion trajectory can be extrapolated back to the beam axis (dashed black). A Higgs is produced and decay promptly into an scalar LLP, that decays at (C) giving rise to a displaced vertex. If the displaced vertex or the pion are at a distance larger than $r_min$ from the primary vertex, the final states is detected as displaced. }}
\end{figure}

At the $pp$ colliders the compressed Higgsinos can be tested by mono-jet plus soft-lepton searches (see e.g~\cite{Schwaller:2013baa,Low:2014cba}) or with disappearing tracks, where the pion from the chargino decay is lost in the hadronic noise of the LHC. At $e^-p$ colliders, Higgsinos will be produced via vector-boson fusion, as shown in the left panel of figure~\ref{fig:1}. Being a $2 \to 4$ process, there is a large phase space suppression as can be seen from the middle panel of figure~\ref{fig:1}, where we compare three $e^-p$ collider setups. We consider first the LHeC~ \cite{AbelleiraFernandez:2012cc}, an add-on using the LHC proton beam of 7 TeV and a 50 GeV electron beam. We then consider the FCC-eh, where we assume a 50 TeV proton beam, and either a 60 GeV~\cite{Zimmermann:2014qxa} or a 240 GeV electron beam. The latter option is not very realistic, and it was considered only as the \emph{blue-sky} setup that would ultimately test the 1.1 TeV thermal Higgsino case.

At $e^-p$ colliders, we will see that we can actually look for the pions, as sketched in the right panel of figure~\ref{fig:1}. The forward jet from the event marks the primary vertex of the collision. The chargino decays into a pion and a neutralino, which we do not show. Unlike the $pp$ case, where one follows the neutralino, here we concentrate on the pion trajectory, and extrapolate it back to the beam axis. If the pion seems to be coming from a point at a distance $r > r_{min}$ and has a transverse momenta $p_T > p_T^{\rm track}$ then we have a \emph{displaced pion}. For the Higgs signature (to be explained in the next section) we have a similar case, except that we directly tag a displaced vertex.

We use here $p_T^{\rm track}=100$ MeV and  $r_{\rm min} = 40 \mu m$ which corresponds to 5 nominal detector resolutions. In ref~\cite{Curtin:2017bxr} we have explored that having $p_T^{\rm track}=50, 400$ MeV and $r_{\rm min} = 25,80 \mu m$ does not affect our conclusions. We always take for simplicity a 100 \% efficiency for the reconstruction of the charged track. 

\begin{figure}[tp]
\begin{center}
\includegraphics[width=4.7cm]{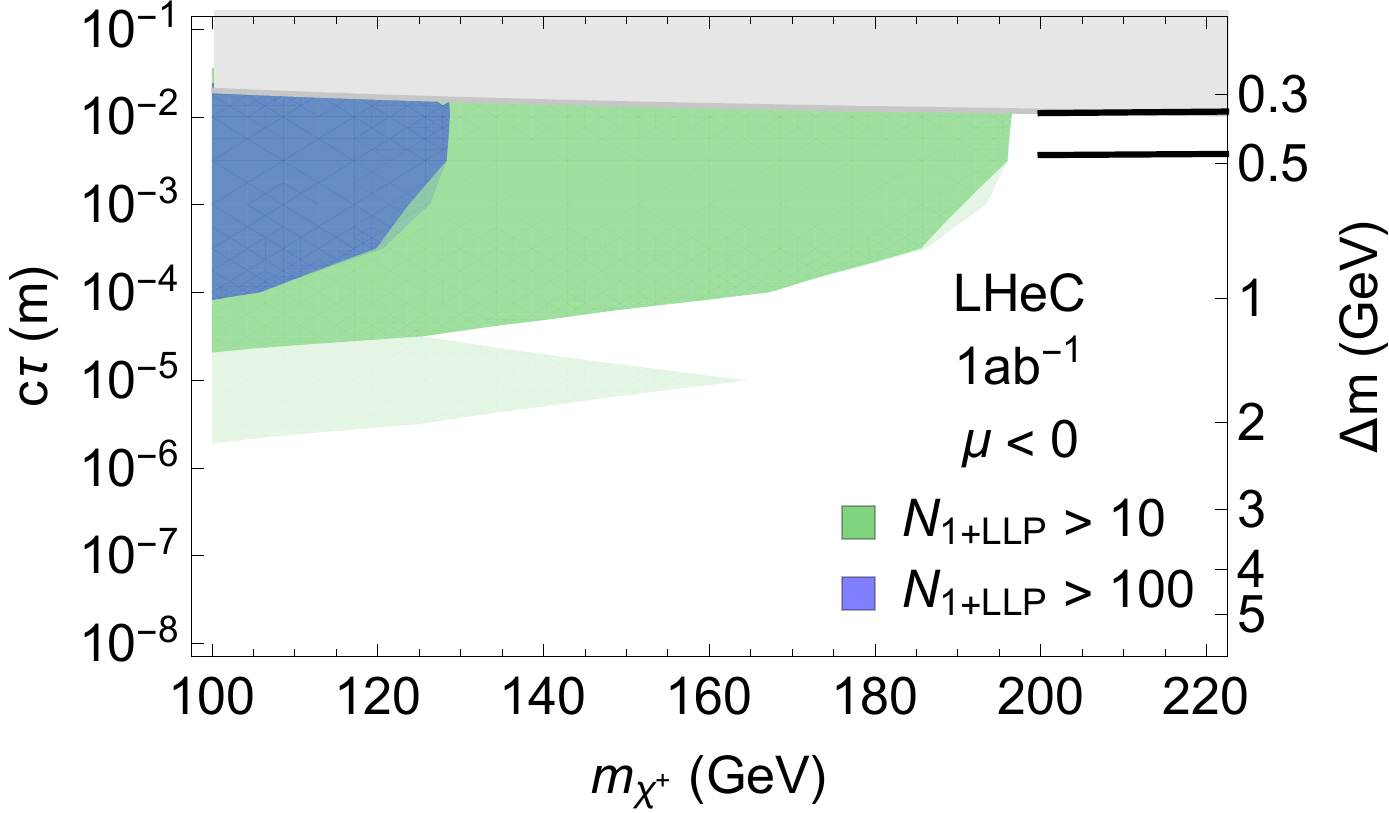}
\hspace{2mm}
\includegraphics[width=4.7cm]{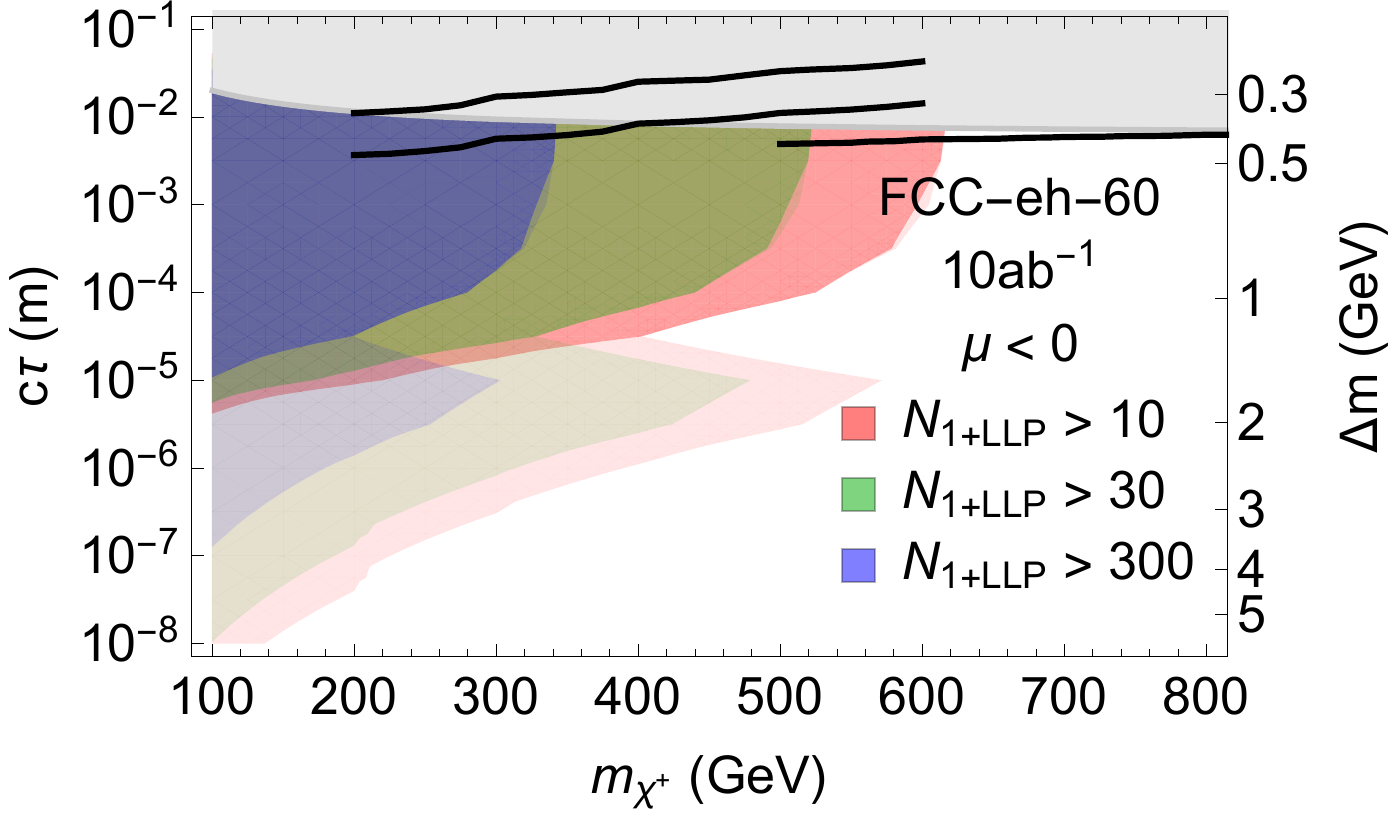}
\hspace{2mm}
\includegraphics[width=4.7cm]{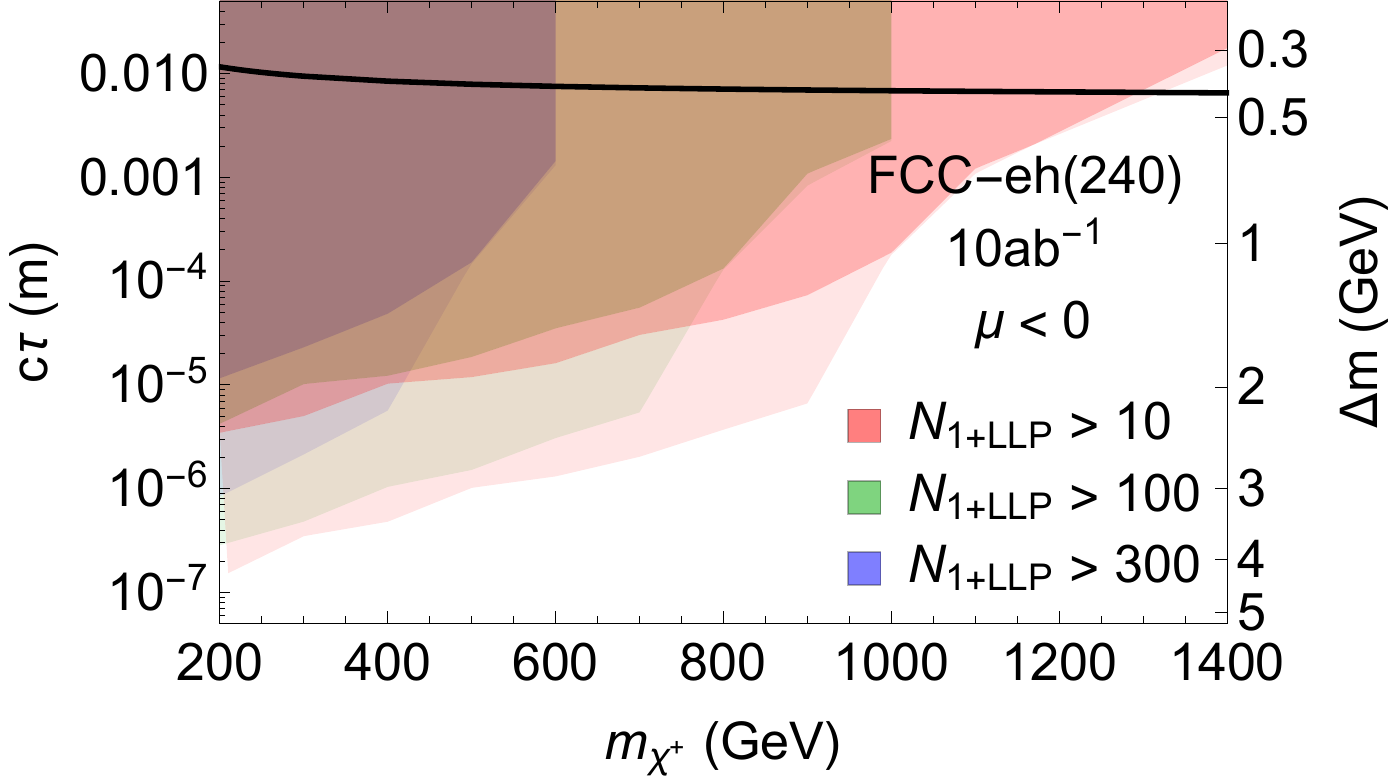}
\end{center}
\vspace{-0.5cm} 
\caption{\label{fig:2}{Reach of the displaced pion search for the different $e^-p$ collider configurations (see main text for details). The grey bands correspond to larger lifetimes than those present in the Higgsino model, and the parallel black lines correspond to the reach of the disappearing track searches at HL-LHC with 3 ab$^{-1}$, taken from the phenomenological study of ref~\cite{Mahbubani:2017gjh}. In the right panel, for clarity reasons, we have removed the disappearing track reach and also traded the grey region by a solid black line showing the $c \tau$ of the pure Higgsino scenario.  }}
\end{figure}

The results of our analysis are shown in figure~\ref{fig:2}, where we show contours of different number of signal events for the $e^-p$ colliders under consideration. Since a large component of the background for these searches would be of instrumental nature and requires an \emph{in-situ} determination, as done for the LHC disappearing track searches~\cite{Aaboud:2017mpt,Sirunyan:2018ldc}, we instead prefer to show contours for 10 and 100 signal events at the LHeC, and 10, 100 and 300 events at the FCC-eh. Here we only present the results for events with one displaced pion, but in ref~\cite{Curtin:2017bxr} the results for two displaced pions were also considered.

In the left panel we see the case of the LHeC with a total integrated luminosity of 1 ab$^{-1}$, which can reach about 180 GeV in mass. This is a bit below the 250 GeV mass reach of the LHC for those Higgsinos~\cite{Schwaller:2013baa}, but we can see that the LHeC is able to probe lifetimes down to 1 mm for most of this mass range, while the LHC is currently struggling to test lifetimes of about 1 cm with disappearing tracks~\cite{Aaboud:2017mpt,Sirunyan:2018ldc}. In the central panel we see the case for the FCC-eh with 10 ab$^{-1}$ of luminosity, and we see the mass reach is close to 600 GeV, which is also what a mono-jet analysis is expected to reach~\cite{Low:2014cba}, while the reach in lifetime has moved down by about one-two orders of magnitude. Finally, the right panel correspond to an over-optimistic scenario, that was designed precisely to reach 1.1 TeV Higgsinos with a nominal lifetime.

\section{Exotic Higgs decays at electron-proton colliders}

The current data on the 125 GeV Higgs boson allows for a non-standard (exotic) branching fraction of about 10 \%. Such exotic decays are theoretically very well motivated (see~\cite{Curtin:2013fra} for a review) and appear naturally in the context of Neutral Naturalness (see e.g~\cite{Craig:2015pha}) and Hidden Valley models~\cite{Strassler:2006im,Strassler:2006ri}. These exotic decays can be realised via the Higgs portal operator, $H^2 \phi^2$ where $H$ is the SM-Higgs doublet before electroweak symmetry breaking and $\phi$ is a new scalar. After electroweak symmetry breaking both scalars mix, giving rise to an H-X-X vertex whose strength is suppressed by the mixing angle, thus making X automatically an LLP. We take the regime where $m_{LLP} < m_h/2 $. These exotic decays can be parametrized in terms of two free parameters, tah can be chosen to be the $BR(H \to X X)$ and the nominal lifetime $c \tau$ of the LLP (or conversely, $m_{LLP}$. We note that since the HL-LHC is expected to produce about $10^8$ Higgs bosons then one naively expect to have access to branching ratios even below $10^{-4}, 10^{-5}$.

\begin{figure}[tp]
\begin{center}
\includegraphics[width=15.8cm]{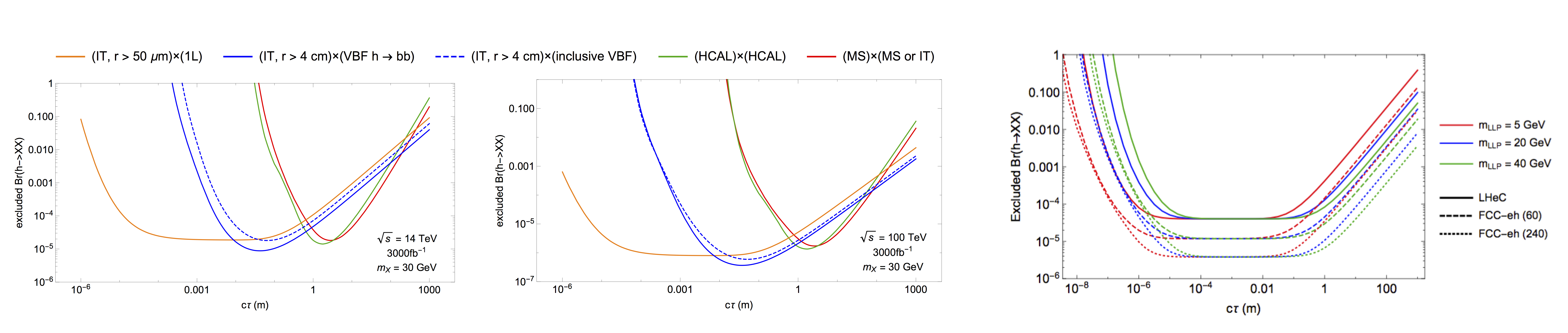}
\end{center}
\vspace{-0.5cm} 
\caption{\label{fig:3}{Reach of the Exotic Higgs decays for the LHC (left), FCC (middle) and the studied $e^-p$ setups (right) in the lifetime-vs-exotic branching fraction plane. See main text for details }}
\end{figure}

We show indeed in figure~\ref{fig:3} the results for the exotic Higgs decays at the LHC (left panel), FCC (center panel) and for our $e^-p$ colliders (right). The first two plots are taken from ref~\cite{Curtin:2015fna}. We indeed see that one can cover exotic branching fractions of $10^{-4}, 10^{-5}$ and compete hand-to-hand with the $pp$ colliders. Moreover, the $e^-p$ colliders provide an advantage for lower lifetimes, reaching all the way down to $c \tau \sim 1 \mu m$. We stress again that the good coverage in lifetime is due to the unique clean environment of these machines.

\section{Conclusions}

In this talk I have discussed how $e^-$p colliders are uniquely suited to look for New Physics by focusing on \emph{hadronic noise} signatures of p-p colliders. A nice example of such example is provided by LLPs, which are omnipresent in models trying to solve fundamental problems of the SM, and notoriously experimentally challenging at hadron colliders. Here I have shown two examples for LLPs at  $e^-$p colliders: (a) Higgsino DM, which gives rise to \emph{displaced pions} is a novel signature which is absolutely impossible to do at the LHC and (b) Exotic Higgs decays, where we find the prospects to be competitive with the LHC reach, testing down to $c \tau$ of a millimeter and branching ratios as low as $10^{-6}$.

\end{document}